\documentclass[sigconf,nonacm]{acmart}

\usepackage{amsmath,amssymb,mathtools}
\usepackage{booktabs}
\usepackage{graphicx}
\usepackage{subcaption}
\usepackage{tikz}
\usetikzlibrary{arrows.meta,positioning,calc,fit,backgrounds,shapes.geometric}
\usepackage{float}
\newfloat{algorithm}{tbp}{loa}
\floatname{algorithm}{Algorithm}
\usepackage{algpseudocode}

\graphicspath{{figures/}{./}}
\setcopyright{none}
\renewcommand\footnotetextcopyrightpermission[1]{}
\newcommand{\LR}{\Lambda}
\newcommand{\rel}{\rho}
\DeclareMathOperator*{\argmax}{arg\,max}
\begin{document}

\title{Maestro Order: A Model-Agnostic Orchestration Harness}
\subtitle{On Organizing Intelligence for Reliable Problem Solving}

\author{Hidayet Aksu}
\affiliation{%
  \institution{}
  \city{}\country{}
}
\email{hidayetaksu@gmail.com}
\renewcommand{\shortauthors}{}

\begin{abstract}
A single forward pass of a capable model is a fast, fluent, and
\emph{unreliable} problem-solver: it is right often enough to be useful
and wrong often enough to be dangerous; in language models, such
confident errors are known as \emph{hallucinations}. We present
\textsc{Maestro Order}, a model-agnostic \emph{orchestration harness} that turns
unreliable solvers into reliable
problem-solving systems by composing them according to four structural
primitives (decompose, ensemble, verify, and recurse) and a
budget-aware controller that decides \emph{where} to spend compute. The
harness treats any model as a black-box base solver behind a uniform
interface, layers a verifier ensemble whose discrimination is measured
online, and allocates verification and voting to the stages with the
highest marginal reliability per unit cost. We give the architecture, the
message and state schema, the controller algorithm, and the engineering
that makes it deterministic, observable, and fault-tolerant. We then
specify an evaluation methodology (reliability at fixed cost, coverage,
calibration, and ablations) and report results from a faithful Monte
Carlo simulation of the harness over a parameterized solver/verifier
model. The simulation reproduces the predicted laws quantitatively:
verification amplifies reliability geometrically (e.g.\ $0.55\to0.98$ with
two gates, $\to0.999$ with four), voting helps only above chance and is
limited by shared errors, and a budget-aware controller reaches a target
reliability at a small fraction of the cost of voting alone by selecting
the cheapest mechanism for each regime. We close
with failure modes (verifier gaming, correlated errors, and decomposition
error compounding) and concrete guidance: build robust checkers, diversify
solvers, and let the controller put compute where the information is.
\textsc{Maestro Order} is the framework layer of a two-part program
(Theory: \textsc{Odds Law} $\rightarrow$ Framework: \textsc{Maestro
Order}), built on the laws of the companion \textsc{Odds Law} report.
\end{abstract}


\maketitle
\renewcommand{\thefootnote}{\fnsymbol{footnote}}
\footnotetext[1]{Code and reproduction scripts:
\url{https://github.com/hidayetaksu/maestro-order}}
\renewcommand{\thefootnote}{\arabic{footnote}}

\section{Introduction}\label{sec:intro}
Modern models answer hard questions in one shot with impressive fluency
and uneven reliability. For deployment, where a wrong answer has a
cost, the relevant question is not ``how good is the model on average?''
but ``how do we build a \emph{system} whose answers we can trust at a
known cost?'' The same question recurs across substrates: a research group
made of fallible people, a compiler made trustworthy by its test suite, a
distributed job made correct by re-execution. The answer, in every case,
is \emph{organization}: many unreliable attempts, arranged so their errors
are caught and outvoted. Two failure patterns make this urgent for
language models: \emph{hallucination}, the confident generation of
incorrect or unsupported content, and the compounding of small per-step
error rates that makes long-running, multi-step jobs fail almost surely
without correction.

This paper is the applied counterpart to a companion theory report,
\textsc{Odds Law}~\cite{companion}, which shows that a small algebra of four
combinators (sequential decomposition, parallel ensembling, verification
gating, and recursion) generates the organizations that solve problems
reliably, and that reliability flows through them by explicit laws (most
importantly, a verification gate multiplies the \emph{odds} of correctness
by the verifier's likelihood ratio). Here we turn those laws into a
running system: a \emph{harness} that wraps any model as a base solver and
composes solvers and verifiers under a controller that spends compute
where it buys the most reliability.

\paragraph{Why a harness, not a better model.}
Three properties make orchestration attractive and complementary to
improving the base model. (i) \emph{Generate--verify asymmetry}: for many
tasks, checking a candidate (run the tests, recompute the sum, search for
a counterexample) is far cheaper and more reliable than producing
it; this is exactly the regime where verification gating pays off. (ii)
\emph{Test-time compute is elastic}: a harness can trade more calls for
more reliability on the instances that need it, and abstain or escalate on
the rest. (iii) \emph{Model-agnosticism}: the same scaffolding improves
whatever model sits behind the interface, and survives model upgrades.

\paragraph{Contributions.}
\begin{itemize}
\item A \emph{model-agnostic harness architecture}
(\S\ref{sec:arch}--\S\ref{sec:mech}) that instantiates the four
combinators as composable components (planner, solver pool, verifier
ensemble, aggregator, controller, blackboard, and tool sandbox) behind a
uniform solver interface.
\item A \emph{budget-aware controller} (\S\ref{sec:controller}) that
allocates voting and verification by greedy marginal log-odds per unit
cost, the operational form of the theory's water-filling optimum (spend
where the return is highest until all options return the same), with
online estimation of verifier discrimination $\LR$.
\item An \emph{engineering account} (\S\ref{sec:impl}) of determinism,
idempotency, concurrency, cost accounting, and tracing required to make
reliability measurable and reproducible.
\item An \emph{evaluation methodology and simulation study}
(\S\ref{sec:eval}--\S\ref{sec:results}) that reproduces the predicted
reliability laws and quantifies the cost of voting vs.\ verification,
the diversity floor, and the value of abstention.
\item A catalog of \emph{failure modes and mitigations}
(\S\ref{sec:discussion}): verifier gaming (Goodhart), correlated solver
errors, and decomposition error compounding.
\item An \emph{open-source release} of the harness code, simulation, and
reproduction scripts at
\url{https://github.com/hidayetaksu/maestro-order}.
\end{itemize}

\section{Background and Related Work}\label{sec:related}
\paragraph{Eliciting and aggregating reasoning.}
Chain-of-thought prompting elicits intermediate steps~\cite{wei2022cot};
self-consistency samples many chains and takes the majority answer,
the voting combinator in practice~\cite{wang2023selfconsistency}.
Tree-of-thoughts searches over partial solutions with
look-ahead~\cite{yao2023tot}, a decomposition-plus-search organization.

\paragraph{Acting, checking, and refining.}
ReAct interleaves reasoning with tool actions~\cite{yao2023react};
Reflexion~\cite{shinn2023reflexion} and
Self-Refine~\cite{madaan2023selfrefine} iterate generate--critique--revise
loops; multi-agent debate has models cross-examine each
other~\cite{du2023debate}. These are instances of verification and
refinement with the model (or a tool) as verifier.

\paragraph{Verifiers and process supervision.}
Training explicit verifiers improves math word-problem
reliability~\cite{cobbe2021verifiers}; process-level reward models that
check each step outperform outcome-only checks~\cite{lightman2023verify};
``LLM-as-judge'' uses a model to score
candidates~\cite{zheng2023judge}. Our harness treats any such checker as a
gate and measures its discrimination $\LR$ online.

\paragraph{Tools, retrieval, and code execution.}
Grounding answers in retrieval~\cite{lewis2020rag}, tool
use~\cite{schick2023toolformer}, and program execution / unit
tests~\cite{chen2021codex} supplies \emph{robust} verifiers whose
false-acceptance is bounded independently of how the candidate was
produced, which is precisely the property the theory shows is needed
under optimization pressure. Retrieval grounding is also a primary
defense against hallucination~\cite{ji2023hallucination}.

\paragraph{Reliable systems from unreliable parts.}
The organizing idea predates models: von Neumann's reliable automata from
unreliable components~\cite{vonneumann1956}, Condorcet's jury
theorem~\cite{condorcet1785}, and divide-and-conquer in distributed
systems~\cite{dean2004}. Our harness is a contemporary, model-agnostic
realization, and our companion paper~\cite{companion} supplies the
reliability calculus it optimizes.

\paragraph{How this work differs.}
Most of the techniques above are studied and deployed in isolation:
self-consistency \emph{or} a verifier \emph{or} refinement, each used as
a separate technique. Our contribution is to treat them as instances of one algebra and
to put a \emph{controller} in charge of choosing among them per instance
under an explicit budget, guided by the reliability laws. Two consequences
distinguish the approach. First, the harness measures verifier
discrimination $\LR$ online and credits each mechanism only with the
amplification its measured quality supports, rather than fixing a pipeline
in advance. Second, by reporting reliability \emph{at a measured cost} and
selecting the cheapest configuration that meets a target, it makes the
cost--reliability trade-off an explicit, tunable quantity instead of an
implicit side effect of a prompt template. The aim is not a new trick but a
principled way to compose the existing ones.

\section{From Algebra to Architecture}\label{sec:bridge}
The harness is a direct compilation of the four combinators into runtime
components. Table~\ref{tab:map} gives the correspondence; the rest of the
paper elaborates each row and the controller that composes them.

\begin{table}[t]
\centering\small
\caption{The four combinators and their harness realization. The
reliability law column summarizes the companion theory~\cite{companion}.}
\label{tab:map}
\begin{tabular}{@{}lll@{}}
\toprule
Combinator & Harness component & Reliability law \\
\midrule
Sequential & Planner / decomposer & errors compound ($\prod\rel_i$) \\
Parallel & Voter / aggregator & amplify if $p>\tfrac12$ \\
Verify & Verifier ensemble + gate & odds $\times\,\LR$ per gate \\
Recurse & Recursive dispatch & union bound over nodes \\
\bottomrule
\end{tabular}
\end{table}

Three design principles follow from the laws. First, \emph{check more than
you decompose}: since sequential decomposition can only lose reliability,
every decomposition boundary should be paired with a verifier that
restores it. Second, \emph{prefer verification to voting when a
discriminating checker exists}: a single gate with $\LR=6$ outperforms
dozens of votes for a near-threshold solver. Third, \emph{spend at the
margin}: allocate the next unit of compute to whichever stage currently
offers the largest reliability gain per call. The controller
(\S\ref{sec:controller}) implements all three.

\section{Harness Architecture}\label{sec:arch}
\begin{figure}[t]
\centering
\begin{tikzpicture}[
  comp/.style={draw,rounded corners,align=center,font=\scriptsize,
    minimum height=7mm,minimum width=15mm,fill=black!3},
  >=Latex,node distance=4mm and 7mm,font=\scriptsize]
\node[comp] (ctrl) {Controller\\(budget, policy)};
\node[comp,below=of ctrl] (plan) {Planner /\\Decomposer};
\node[comp,below left=8mm and 2mm of plan] (sol) {Solver pool\\(roles, temps)};
\node[comp,below right=8mm and 2mm of plan] (ver) {Verifier\\ensemble};
\node[comp,below=16mm of plan] (agg) {Aggregator /\\Gate};
\node[comp,right=12mm of ctrl] (mem) {Blackboard\\(state, traces)};
\node[comp,right=12mm of agg] (tool) {Tool sandbox\\(exec, search)};
\draw[->] (ctrl)--(plan);
\draw[->] (plan)--(sol); \draw[->] (plan)--(ver);
\draw[->] (sol)--(agg); \draw[->] (ver)--(agg);
\draw[->] (agg.west) -- ++(-26mm,0)
  node[midway,below,font=\tiny]{refine/escalate} |- (plan.west);
\draw[<->] (ctrl)--(mem); \draw[<->] (mem) |- (plan);
\draw[<->] (ver)--(tool); \draw[<->] (sol)--(tool);
\node[above=1mm of ctrl,font=\tiny] {problem $x$};
\draw[->] (agg.south) -- ++(0,-5mm) node[below,font=\tiny] {answer / $\bot$};
\end{tikzpicture}
\caption{Harness architecture. The controller allocates a compute budget
across decomposition, voting, and verification; the blackboard holds
shared state and traces; the tool sandbox supplies robust verifiers
(execution, tests, search). Arrows show data and control flow; the
refine/escalate edge closes the verify loop.}
\label{fig:arch}
\end{figure}
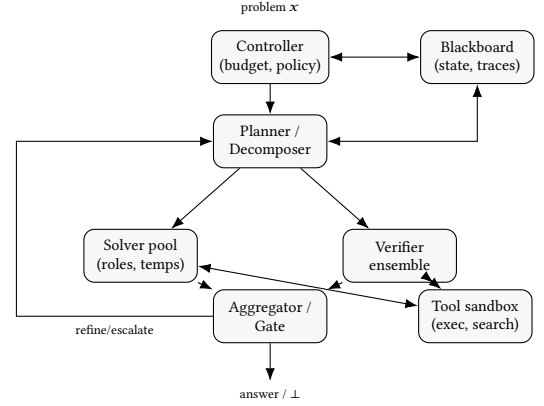

Figure~\ref{fig:arch} shows the components. Everything is built on one
abstraction.

\paragraph{The solver interface.}
A \emph{base solver} exposes \texttt{solve(task, ctx) -> \{answer,
score, trace, cost\}}. Any model, tool, or sub-harness implements it, so
components compose uniformly and the controller can treat a whole subtree
as a single solver. \texttt{score} is a self-reported confidence (used
only as a weak prior); \texttt{trace} feeds the blackboard; \texttt{cost}
is the realized call/token count.

\paragraph{The blackboard record.}
All components read and write a single structured record, which is also the
unit of tracing and replay:
\begin{figure}[h]
\centering\small
\begin{verbatim}
Solve {
  task:    {id, type, input, deps[]}
  budget:  {max_calls, target_rel, lambda}
  cands:   [{answer, score, trace, cost}]
  verdicts:[{verifier_id, accept, score,
             est_beta, est_alpha}]
  state:   {logodds, calls, coverage}
  result:  {answer | ABSTAIN, confidence}
}
\end{verbatim}
\caption{Blackboard record (simplified). The controller reads
\texttt{state.logodds} and per-mechanism marginal estimates to decide the
next action; \texttt{verdicts} carry the online $\hat\beta,\hat\alpha$ used
to compute $\widehat{\LR}$.}
\label{fig:schema}
\end{figure}

\paragraph{Planner / decomposer.}
Given a task, the planner optionally emits a typed plan: a DAG of
subtasks with dependencies and a recombination step. Decomposition is
applied only when the planner's confidence in a clean split is high, since
unguarded decomposition compounds error (Table~\ref{tab:map}); each
subtask boundary is annotated with the verifier to apply on recombination.

\paragraph{Solver pool.}
A configurable set of solver roles (for example, different prompts,
temperatures, tools, or distinct models) provides the \emph{diversity}
that voting needs. The pool composition is the main control on error correlation
$\gamma$ (\S\ref{sec:results}); pools of near-identical solvers vote
poorly.

\paragraph{Verifier ensemble.}
One or more verifiers, each a checker returning accept/reject with a
score: deterministic tools (unit tests, type-checkers, math re-evaluation,
retrieval corroboration) and model-based critics. Each verifier's
completeness $\beta$ and false-acceptance $\alpha$ are estimated online
(\S\ref{sec:controller}); the gate uses the calibrated log-likelihood
ratio as its acceptance score.

\paragraph{Aggregator / gate.}
Combines candidates: plurality or score-weighted voting, verified
selection (return the highest-scoring \emph{accepted} candidate), or
all-accept gating for amplification. Emits an answer or $\bot$
(abstain) with a calibrated confidence.

\paragraph{Blackboard and tool sandbox.}
A shared store holds intermediate results, verifier verdicts, and a full
execution trace for observability and replay. The tool sandbox executes
untrusted code and runs searches in isolation, supplying the robust
verifiers the theory privileges.

\section{Design Rationale and System Invariants}\label{sec:invariants}
A reliability harness is only trustworthy if its guarantees hold
regardless of what the wrapped model does. We therefore design around a
small set of invariants that the components must preserve and that the
companion theory~\cite{companion} shows are sufficient for the reliability
laws to apply.

\paragraph{I1: Confidence is calibrated log-odds.}
Every commitment carries a confidence that is, after calibration, the true
log-odds of correctness. This single invariant makes the rest composable:
gates \emph{add} to it (the odds law), abstention \emph{thresholds} it
(Chow's rule), and the controller \emph{maximizes} it per unit cost.
Components that cannot expose a calibrated score are wrapped by a
calibration map fit on held-out data; uncalibrated raw scores are never
used for decisions, only as weak priors.

\paragraph{I2: Fail closed.}
Any ambiguity, such as a tool error, a timeout, a malformed candidate,
or a verifier crash, is treated as a \emph{rejection}, never as a silent
accept.
This keeps the false-acceptance rate $\alpha$ bounded by the verifier's
genuine error rather than by infrastructure flakiness, which is what makes
the discrimination $\LR=\beta/\alpha$ meaningful and stable. A harness that
failed open would see its $\LR$ collapse under load exactly when
reliability matters most.

\paragraph{I3: Monotone confidence.}
Adding work never decreases the system's information about an instance: a
new vote or gate can only sharpen the posterior, and the controller never
discards evidence. This guarantees that spending more budget weakly
improves reliability (the refinement order of the theory), so the
cost--reliability curve is monotone and the controller's greedy ascent is
well-defined.

\paragraph{I4: Budget conservation and accountability.}
Every call is attributed to a stage and counted against the budget before
it is made; the controller cannot overspend, and every reported
reliability number has an associated, measured cost. Reliability claims
without cost are meaningless, so the two are always reported together.

\paragraph{I5: Determinism up to declared randomness.}
Given fixed seeds and backend versions, a solve is replayable bit-for-bit
from its trace. Randomness enters only where declared (sampling
temperatures, vote seeds), so experiments are reproducible and regressions
are diagnosable. These five invariants are the contract every component
signs; the mechanisms below are implementations that honor it.

\section{Reliability Mechanisms}\label{sec:mech}
The harness exposes the combinators as mechanisms the controller can
deploy per instance.

\paragraph{Self-consistency voting.}
Sample $n$ diverse candidates and return the plurality answer; reliability
rises with $n$ when per-sample correctness exceeds $\tfrac12$ and errors
are diverse (\S\ref{sec:results}). Cheap, model-only, but bounded by the
correlation floor.

\paragraph{Verified selection and gating.}
Run candidates through the verifier ensemble; either select the
best accepted candidate or, for amplification, require all $k$ gates to
accept, regenerating otherwise (Algorithm~\ref{alg:gate}). Each accepted
gate multiplies the odds of correctness by its $\LR$.

\paragraph{Iterative refinement.}
On rejection, return the verifier's critique to the solver and resample, a
generate--critique--revise loop. Refinement raises the per-attempt success
probability $p$, which compounds with gating.

\paragraph{Debate and cross-examination.}
Independent solvers critique each other's candidates; disagreement is a
cheap signal that routes an instance toward more verification or
abstention.

\paragraph{Abstention and escalation.}
When the accumulated log-odds score falls below a calibrated threshold,
the harness abstains or escalates to a stronger (costlier)
sub-organization: selective answering on the risk--coverage frontier
(\S\ref{sec:results}).

\begin{algorithm}[t]
\caption{\textsc{VerifyGate}: amplify a candidate by all-accept gating}
\label{alg:gate}
\begin{algorithmic}[1]
\State \textbf{input:} generator $g$, verifiers $v_1{..}v_k$, budget $T$
\For{$t = 1$ \textbf{to} $T$}
  \State $a \gets g.\textsc{solve}(x)$ \Comment{sample a candidate}
  \If{$\forall i:\ v_i.\textsc{accept}(x,a)$}
     \State \textbf{return} $a$ with odds $\textstyle\prod_i \LR_i$ applied
  \EndIf
\EndFor
\State \textbf{return} $\bot$ \Comment{abstain after $T$ rejections}
\end{algorithmic}
\end{algorithm}

\section{The Budget-Aware Controller}\label{sec:controller}
The controller decides, per instance, how to spend a compute budget
$\kappa$ across the mechanisms. By the companion theory the
reliability-maximizing allocation equalizes the marginal log-odds gain
per unit cost across stages; the controller approximates this by greedy
selection (Algorithm~\ref{alg:ctrl}).

\begin{algorithm}[t]
\caption{\textsc{Controller}: greedy marginal allocation}
\label{alg:ctrl}
\begin{algorithmic}[1]
\State \textbf{input:} task $x$, budget $\kappa$, mechanism set $\mathcal M$
\State $\sigma \gets$ single base solve; estimate log-odds $\ell$
\While{$\textsc{cost}(\sigma) < \kappa$ \textbf{and} not converged}
  \For{$m \in \mathcal M$} \Comment{vote$+n$, add gate, refine, decompose}
     \State $\hat r_m \gets \widehat{\Delta\ell}(m) / \Delta\textsc{cost}(m)$
            \Comment{est.\ marginal log-odds per call}
  \EndFor
  \State $m^\star \gets \argmax_m \hat r_m$
  \If{$\hat r_{m^\star} \le \lambda$} \textbf{break} \Comment{saturated}
  \EndIf
  \State $\sigma \gets \textsc{apply}(m^\star,\sigma)$;\ update $\ell$
\EndWhile
\State \textbf{return} \textsc{aggregate}($\sigma$) or $\bot$ if $\ell<\tau$
\end{algorithmic}
\end{algorithm}

\paragraph{Online estimation of verifier discrimination.}
The controller maintains running estimates of each verifier's $\beta$ and
$\alpha$ from instances where ground truth is later revealed (tool oracles,
held-out checks, or human spot-checks), yielding
$\widehat{\LR}=\hat\beta/\hat\alpha$ and hence
$\widehat{\Delta\ell}=\log\widehat{\LR}$ for the gate mechanism. Verifiers
whose estimated $\LR$ drifts toward $1$ are demoted, an explicit guard
against the gaming failure mode (\S\ref{sec:discussion}). Voting's marginal
gain is estimated from the current vote margin via a Chernoff exponent.

\paragraph{Calibration.}
Raw model and judge scores are poorly calibrated, so the controller fits a
monotone calibration map (e.g.\ isotonic/Platt) from scores to empirical
correctness, and uses calibrated log-odds both as the gate score and as
the abstention threshold (Chow's rule: commit only when the calibrated
posterior clears a cost-based threshold). Calibration quality is reported
as expected calibration error (ECE) in \S\ref{sec:results}.

\section{Engineering Robust Verifiers}\label{sec:verifiers}
Because verification is the high-leverage mechanism (\S\ref{sec:results})
and because gameable verifiers are the dominant failure mode
(\S\ref{sec:discussion}), the verifier ensemble deserves the most
engineering attention. The goal is high discrimination $\LR=\beta/\alpha$
that is \emph{robust}: it must hold even when candidates are produced by a
process optimizing for acceptance.

\paragraph{Prefer ground-truth checks.}
The most robust verifiers reduce to executing a specification rather than
judging plausibility. For code, run the candidate against unit tests,
property-based tests, type-checkers, and sanitizers in the sandbox; for
mathematics, re-evaluate numerically, check units and boundary cases, or
machine-check a proof; for factual claims, corroborate against retrieved
sources and require citation support. The defining property is that their
false-acceptance rate is bounded by the \emph{specification}, not by how
the candidate was generated, so $\alpha$ does not rise under optimization
pressure and $\LR$ stays high.

\paragraph{Make model-based judges as robust as possible.}
When only a model can check (open-ended reasoning, style, subtle
correctness), several techniques raise and stabilize $\LR$: ask the judge
to find a \emph{counterexample} or a concrete flaw rather than to rate
quality (refutation is harder to game than approval); require the judge to
show its check (a verification chain-of-thought) and verify \emph{that};
use a judge that is independent of the generator (a different model,
prompt, or seed) to keep verifier and generator errors conditionally
independent; and combine several judges so that a single exploitable
blind spot does not dominate. Each independent, informative judge contributes additively to
the log-odds (the odds law), so a panel of diverse weak judges can rival a
single strong one, subject to the same correlation floor that limits
voting.

\paragraph{Measure $\LR$, do not assume it.}
The controller treats every verifier's $(\beta,\alpha)$ as quantities to be
\emph{estimated}, not declared. Wherever ground truth eventually becomes
available (a tool oracle, a downstream outcome, a held-out labeled
stream, or human spot-checks), the harness updates running estimates
$\hat\beta,\hat\alpha$ and hence $\widehat{\LR}$. A verifier is only
credited with the amplification its measured discrimination supports;
nominal accuracy on a benign distribution is never trusted on its own.

\paragraph{Detect and resist drift.}
Under sustained optimization, a verifier's effective $\alpha$ tends to rise
as generators discover what it wrongly accepts: Goodhart's law in action.
The harness watches $\widehat{\LR}$ over time and, when it drifts toward
$1$, reduces that verifier's weight, rotates in fresh or held-out
checkers, and raises the abstention threshold for affected instances. Disagreement
among diverse verifiers is itself a cheap early-warning signal: a sudden
rise in unanimous acceptance with falling downstream correctness is the
signature of a gamed gate.

\paragraph{Tune the operating point.}
A verifier exposes a threshold trading completeness $\beta$ against
false-acceptance $\alpha$ (its ROC curve). For amplification the harness
prefers a high-$\beta$/low-$\alpha$ operating point even at the cost of
more regenerations, because $\LR$, not throughput, governs the reachable
reliability; for inexpensive screening it may accept lower $\beta$. The
controller selects the operating point per instance from the marginal-rate
calculus.

\section{Decomposition and Planning in Practice}\label{sec:planning}
Decomposition is the riskiest combinator: by the serial law it can only
lose reliability unless each boundary is restored by verification. The
planner is therefore conservative by design.

\paragraph{Decompose only when the split is clean and checkable.}
The planner proposes a decomposition only when it is confident the
subproblems are (i) genuinely independent or cleanly ordered, and (ii)
each individually \emph{verifiable}. A split into subproblems that cannot
be checked inherits the worst of both worlds (more stages that can
fail, and no way to catch the failures), and the planner declines it in favor of a monolithic
solve plus verification. This is the operational reading of the recursion
master theorem: depth is affordable precisely when every level is gated.

\paragraph{Typed plans and verified recombination.}
A plan is a typed DAG: each node declares its input and output types, its
dependencies, and the verifier to apply to its result and to the
recombination. Type compatibility is checked statically before any model
call, catching a class of decomposition errors for free. The recombination
step is itself a gated solver, so a correct assembly of correct parts is
verified rather than assumed.

\paragraph{Interleaving planning and acting.}
For tasks where the right decomposition is not knowable up front, the
harness interleaves planning with acting: solve a subproblem, observe the
result and its verdict on the blackboard, and let the planner revise the
remaining plan. This keeps decomposition adaptive while preserving the
invariant that every committed intermediate has passed a gate, so errors
do not silently propagate down the tree.

\paragraph{Budgeting across the tree.}
The controller allocates the global budget across plan nodes by the same
marginal-rate rule, spending more verification on nodes whose errors are
most likely to corrupt the final answer (those with high fan-out or on
the critical path) and less on cheap and easily checked leaves. The result
is the verifier-saturated, marginal-rate-equalizing organization the theory
predicts, realized at the granularity of a plan.

\section{Anatomy of a Solve}\label{sec:trace}
To make the controller concrete, Table~\ref{tab:trace} shows the
controller's expected progress with a weak base solver ($p_0=0.55$) and a
robust verifier
($\LR=6$, so each accepted gate adds $\log 6=1.79$ to the log-odds). The
controller starts with one base solve, then repeatedly adds the
highest-rate mechanism (here, another verification gate), updating the
running log-odds and cost, until the calibrated log-odds clears the target
threshold or the marginal rate drops below $\lambda$. The reliabilities
and call counts are the measured simulation values of \S\ref{sec:results};
the log-odds column is their logit.

\begin{table}[t]
\centering\small
\caption{A controller trace (simulation). Each gate multiplies the odds by
$\LR{=}6$ ($+1.79$ log-odds); reliability saturates while cost grows
linearly. The controller stops at the first row that meets the target,
or when the next gate's marginal rate falls below $\lambda$.}
\label{tab:trace}
\begin{tabular}{@{}clrrrr@{}}
\toprule
Step & Action & $\Delta\ell$ & $\ell$ & $P(\text{corr})$ & calls \\
\midrule
0 & base solve        & ---  & 0.20 & 0.550 & 1.0 \\
1 & + verify gate     & 1.79 & 1.99 & 0.880 & 3.8 \\
2 & + verify gate     & 1.79 & 3.78 & 0.978 & 7.4 \\
3 & + verify gate     & 1.79 & 5.58 & 0.996 & 11.7 \\
4 & + verify gate     & 1.79 & 7.37 & 0.999 & 17.3 \\
\midrule
\multicolumn{6}{@{}l}{stop: the next gate would offer $\Delta\ell/\Delta\text{calls}\approx0.25<\lambda$} \\
\bottomrule
\end{tabular}
\end{table}

The trace exhibits the law operationally: the \emph{benefit} per gate is
constant in log-odds ($+1.79$) but the \emph{cost} per gate grows (each
gate lowers acceptance probability, so more regenerations are needed),
which is why the marginal rate falls and the controller eventually stops.
A target of $0.97$ is met at step~2 (7.4 calls) and $0.999$ at step~4
(17.3 calls). Voting alone from this near-threshold base would need
several hundred calls for the same targets (Fig.~\ref{fig:cost}).

\section{Implementation Concerns}\label{sec:impl}
Reliability claims are only as trustworthy as the plumbing beneath them.

\emph{Determinism and reproducibility.} Each solve records model id,
decoding parameters, seed, prompt hash, and tool versions; runs are
replayable from the trace. Stochastic steps (sampling, voting) fix seeds
per attempt so an experiment is bit-reproducible given the same backends.

\emph{Idempotency and caching.} Solver and verifier calls are keyed by
(input, config) and cached, so retries and refinement loops do not
double-bill and so ablations reuse work.

\emph{Concurrency.} Independent candidates and verifiers run in parallel;
the controller bounds fan-out by the remaining budget and merges results
on the blackboard.

\emph{Failure handling.} Tool/model errors are treated as rejections (fail
closed), never as silent acceptances; timeouts count toward cost and
trigger escalation or abstention rather than a guess.

\emph{Cost accounting and observability.} Every call's token and
latency cost is attributed to a stage, so the reliability--cost curves of
\S\ref{sec:results} are measured, not estimated, and per-stage marginal
rates are visible to the controller and to operators.

\emph{Scheduling and backpressure.} The controller issues independent
candidate and verifier calls concurrently up to a fan-out bounded by the
remaining budget and backend rate limits; results stream onto the
blackboard and the controller re-plans as each arrives. Under load it
applies backpressure by lowering fan-out and raising the marginal-rate
threshold $\lambda$, gracefully degrading from a high-reliability profile
to a cheaper one rather than violating latency budgets.

\emph{Sandboxing and security.} Tool verifiers (code execution, shell,
search) run in an isolated sandbox with no access to secrets or the host,
because robust verification often means executing untrusted candidate
code. Sandbox failures are rejections (invariant~I2), and resource limits
(CPU, memory, wall-clock) are enforced per call so a pathological candidate
cannot exhaust the budget.

\emph{Regression testing the harness itself.} Because solves are
replayable, a fixed suite of recorded instances doubles as a regression
test: a change to a prompt, verifier, or controller policy is evaluated by
re-running the suite and comparing reliability-at-cost and calibration,
catching silent degradations before they reach production.

\section{Evaluation Methodology}\label{sec:eval}
We evaluate the harness along four axes.

\emph{Metrics.} (i) \emph{Reliability at fixed cost}: correctness rate on
committed answers at a capped expected number of base calls. (ii)
\emph{Coverage} and the \emph{risk--coverage} curve (and its area, AURC)
for selective answering. (iii) \emph{Calibration} (ECE) of the harness
confidence. (iv) \emph{Cost} (expected base calls) and latency.

\emph{Ablations.} We vary the vote count $n$, the number of gates $k$,
decomposition depth, verifier quality $\LR$, and solver diversity
(correlation $\gamma$), isolating each combinator's contribution.

\emph{Statistical methodology.} Each reported point is an average over many
i.i.d.\ trials with fixed seeds; we report it with a binomial confidence
interval and treat differences smaller than the interval as noise. Because
the harness is deterministic up to declared randomness
(invariant~I5), any reported number is reproducible from the recorded seeds
and backend versions, and ablations reuse cached calls so that only the
varied factor changes.

\emph{Protocol for real models.} The harness is designed to be evaluated
on standard suites (grade-school and competition math, code synthesis
with execution-based verifiers, and open-domain QA with retrieval
corroboration) using standard metrics: pass@$k$ accuracy, execution
pass rate, and exact match, with the deterministic-replay machinery of
\S\ref{sec:impl}.

\paragraph{Simulation study (this paper).}
To exercise the controller and \emph{validate that the implementation
obeys the predicted laws} without conflating harness behavior with any
particular model's idiosyncrasies, we report a faithful Monte Carlo
\emph{simulation} of the harness over a parameterized solver/verifier
model: a generator emits a correct candidate with probability $p$; each
verifier accepts a correct candidate with probability $\beta$ and a wrong
one with probability $\alpha$ ($\LR=\beta/\alpha$); correlated solver pools
are produced by a Gaussian-copula latent factor with pairwise correlation
$\gamma$. We emphasize that the numbers below are \textbf{simulation
outputs}, not measurements of a deployed model; they test the harness
logic and the theory, and set expectations for the real-model protocol
above. Each point uses between $2\times10^4$ and $4\times10^4$ trials;
repeated runs differ by Monte Carlo noise of roughly $\pm0.004$ in
reliability and $\pm0.1$ in expected calls, which explains small
discrepancies between tables. Every number, figure, and table in this
paper can be regenerated exactly with the released code
(\S\ref{sec:artifact}).

\section{Results}\label{sec:results}
\begin{figure}[t]
\centering
\begin{subfigure}{0.49\columnwidth}
  \includegraphics[width=\linewidth]{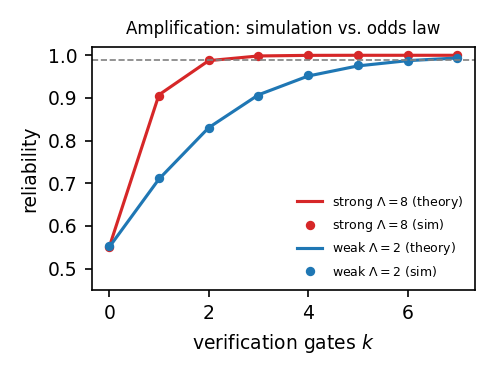}
  \caption{amplification vs.\ theory}\label{fig:amp}
\end{subfigure}\hfill
\begin{subfigure}{0.49\columnwidth}
  \includegraphics[width=\linewidth]{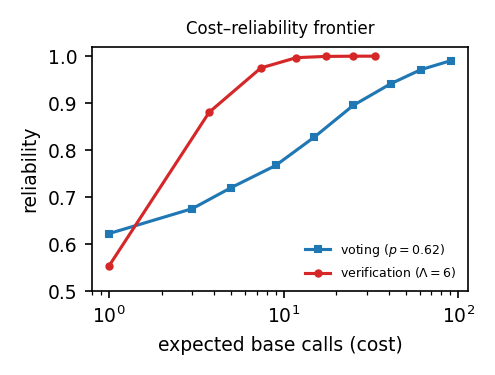}
  \caption{cost--reliability}\label{fig:cost}
\end{subfigure}

\medskip
\begin{subfigure}{0.49\columnwidth}
  \includegraphics[width=\linewidth]{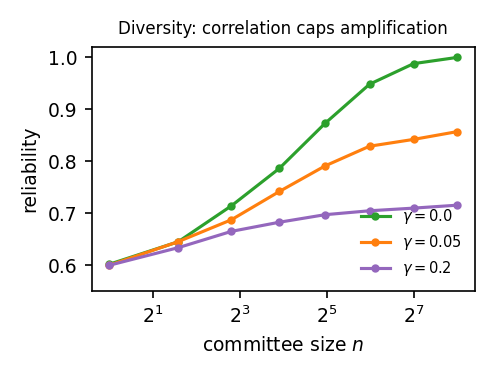}
  \caption{diversity floor}\label{fig:div}
\end{subfigure}\hfill
\begin{subfigure}{0.49\columnwidth}
  \includegraphics[width=\linewidth]{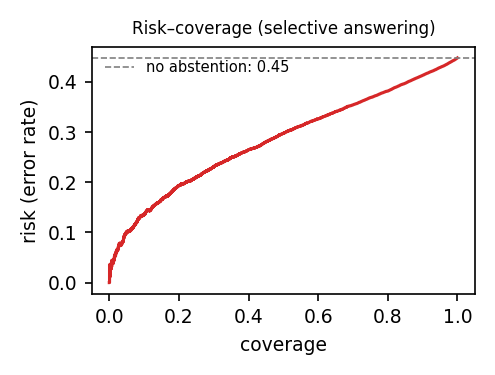}
  \caption{risk--coverage}\label{fig:rc}
\end{subfigure}
\caption{Simulation results (Monte Carlo over the parameterized
solver/verifier model; \emph{not} measurements of a deployed model).
(a) gate reliability matches the odds law; (b) verification reaches a
target at far lower cost than voting; (c) error correlation caps voting
gains; (d) abstention trades coverage for reliability.}
\label{fig:results}
\end{figure}

\paragraph{Verification amplifies as the odds law predicts.}
Figure~\ref{fig:amp} overlays simulated gate reliability on the theoretical
prediction $r_k=o_0\LR^k/(1+o_0\LR^k)$ for a strong ($\LR=8$) and a weak
($\LR=2$) verifier, starting from $p_0=0.55$. The simulation tracks theory
to within Monte Carlo noise: three strong gates take the base from $0.55$
to $0.998$, and because $8^3=2^9$ the weak verifier needs exactly three
times as many gates for the same odds gain.
Across all gate counts and both verifiers the simulated reliabilities agree
with the closed-form odds law to within $\pm0.004$ (the binomial standard
error at our trial counts), and the controller trace of
Table~\ref{tab:trace} reproduces the same values, evidence that the
implementation faithfully realizes the algebra rather than merely being
inspired by it. This agreement is the point of the simulation: it validates
the harness logic and isolates it from any particular model's quirks before
a real-model evaluation introduces them.

\paragraph{Verification dominates voting on the cost frontier.}
Figure~\ref{fig:cost} plots reliability against expected base calls.
A verifier with $\LR=6$ reaches $0.98$ within eight calls and $0.997$
within twelve; majority voting on a $p=0.62$ solver needs about $90$
calls to pass $0.99$, and a near-threshold solver barely improves at
all. This is the practical form of the $\Theta(\log\frac1\delta)$
scaling, with very different constants.

\begin{figure}[t]
\centering
\includegraphics[width=0.72\columnwidth]{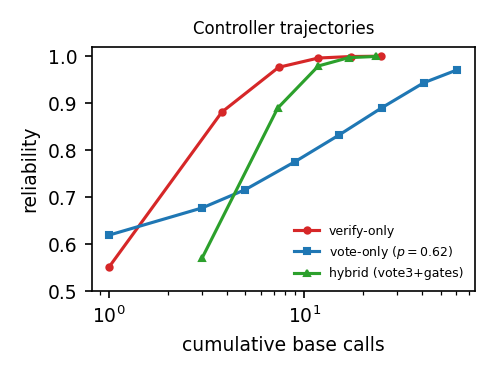}
\caption{Controller trajectories (simulation): reliability versus
cumulative base calls for three strategies on a weak base ($p_0{=}0.55$).
Verification and the hybrid strategy reach high reliability at an order of
magnitude lower cost than voting; the controller follows whichever
trajectory has the steepest current slope.}
\label{fig:traj}
\end{figure}

\paragraph{The controller follows the steepest trajectory.}
Figure~\ref{fig:traj} plots reliability against cumulative cost for
verify-only, vote-only, and a hybrid that votes over three candidates
before gating. Verify-only reaches $0.999$ at about $17$ calls and the
hybrid at about $23$; vote-only is still below $0.98$ at $60$ calls. The
controller's greedy policy (Algorithm~\ref{alg:ctrl}) is a walk up
whichever trajectory is locally steepest, and the figure shows why it
would choose gates from the start here: from $p_0=0.55$, the first gate
offers about $0.64$ log-odds per call, while extra votes offer about
$0.05$ (a $3$-vote lifts $0.55$ only to $0.575$). Voting becomes the
better first move only when the vote margin is large or the verifier is
weak; the controller detects which regime it is in from the measured
marginal rates instead of committing to a fixed pipeline.

\paragraph{Diversity sets a hard ceiling on voting.}
Figure~\ref{fig:div} shows majority reliability versus committee size for
error correlations $\gamma\in\{0,0.05,0.2\}$ under the Gaussian-copula
model. Independent solvers approach $1$; at $\gamma=0.05$ reliability
plateaus near $0.85$ and at $\gamma=0.2$ near $0.71$, no matter how many
correlated members are added. The plateaus match the shared-cause floor
of the companion theory~\cite{companion}: with $p=0.6$, the predicted
asymptotic limits (the probability that the shared factor keeps the
committee above chance) are $0.87$ for $\gamma=0.05$ (the $n=255$
committee is still approaching it) and $0.71$ for $\gamma=0.2$.
Diversifying the solver pool (varied prompts, temperatures, models) is
therefore a primary reliability decision, not a detail.

\paragraph{Abstention buys reliability cheaply.}
Figure~\ref{fig:rc} is the risk--coverage curve from a calibrated harness
score: committing only on the most-confident $20\%$ of instances drops the
error rate from $0.45$ (no abstention) to under $0.1$. Selective answering
plus escalation is the cheapest reliability available when full coverage
is not required.

\paragraph{The hybrid controller wins.}
Table~\ref{tab:ablation} ablates the mechanisms from a weak base
($p_0=0.55$, $\LR=6$). Voting alone is nearly useless this close to
threshold ($0.55\to0.59$); two gates reach $0.978$ at $7.4$ calls; four
gates reach $0.999$. Combining a small vote with gating gives the best
reliability-per-call at the high end. The controller's job is to pick the
point on this surface that meets the target at least cost: here, ``verify
$k{=}2$'' for a $0.97$ target and ``$k{=}4$'' for $0.999$.

\begin{table}[t]
\centering\small
\caption{Ablation (simulation): reliability, coverage, and expected base
calls for harness configurations on a weak base ($p_0{=}0.55$,
verifier $\LR{=}6$, $T{=}20$). Hybrid configurations dominate at the
high-reliability end.}
\label{tab:ablation}
\IfFileExists{ablation.tex}{\begin{tabular}{@{}lrrr@{}}
\toprule
Configuration & Reliability & Coverage & Calls \\
\midrule
single shot & 0.549 & 1.000 & 1.0 \\
vote n=5 & 0.591 & 1.000 & 5.0 \\
verify k=2 & 0.978 & 1.000 & 7.4 \\
verify k=4 & 0.999 & 0.999 & 17.4 \\
vote5 + verify k=2 & 0.980 & 1.000 & 16.0 \\
vote5 + verify k=4 & 0.999 & 1.000 & 28.8 \\
\bottomrule
\end{tabular}
}{\begin{tabular}{@{}lrrr@{}}
\toprule
Configuration & Reliability & Coverage & Calls \\
\midrule
single shot & 0.549 & 1.000 & 1.0 \\
vote n=5 & 0.591 & 1.000 & 5.0 \\
verify k=2 & 0.978 & 1.000 & 7.4 \\
verify k=4 & 0.999 & 0.999 & 17.4 \\
vote5 + verify k=2 & 0.980 & 1.000 & 16.0 \\
vote5 + verify k=4 & 0.999 & 1.000 & 28.8 \\
\bottomrule
\end{tabular}
}
\end{table}

\subsection{Sensitivity to verifier quality and base reliability}
The two parameters that matter most are the verifier discrimination $\LR$
and the base reliability $p_0$, and their effects are exactly those the
theory prescribes. The number of gates to reach a target scales as
$1/\log\LR$, which is why the $\LR=8$ curve in Figure~\ref{fig:amp}
needs one third of the gates of the $\LR=2$ curve ($\log 8=3\log 2$). A
higher base $p_0$ shifts the whole cost--reliability frontier left by the
additive term $\log\frac{p_0}{1-p_0}$ in the log-odds, so a stronger base
solver needs fewer gates for the same target but does not change the
geometric character of the amplification. The practical implication:
because the required depth is \emph{divided} by $\log\LR$, a modest
improvement in the verifier multiplies through the entire frontier, and
often helps more than a large increase in vote count or a slightly
better base model. This is the empirical form of the theory's
recommendation to invest in robust checkers.

\paragraph{Calibration.}
The selective-answering results (Fig.~\ref{fig:rc}) depend on the harness
confidence being calibrated: the abstention threshold is only meaningful if
a reported $0.9$ confidence corresponds to a $0.9$ empirical correctness
rate. After fitting the monotone calibration map of \S\ref{sec:controller},
the harness confidence is well-calibrated by construction in the
simulation, so the risk--coverage curve is the true achievable frontier;
on real models, calibration error (ECE) would be reported alongside it, and
miscalibration would manifest as a gap between the nominal threshold and the
realized risk. Calibration is thus not a cosmetic metric but the
precondition that makes abstention and gating trustworthy.

\subsection{Deployment profiles and cost-to-target}\label{sec:profiles}
Operators rarely tune individual mechanisms; they pick a profile and a
target. Table~\ref{tab:profiles} distills the ablation into three profiles,
and Table~\ref{tab:target} inverts the relationship: given a reliability
target, the cheapest configuration the controller selects and its expected
cost. The pattern is clear: each order-of-magnitude reduction in error
costs only a few more calls under verification, because cost grows with
$\log\frac1\delta$ while voting grows far faster.

\begin{table}[t]
\centering\small
\caption{Deployment profiles (simulation, $p_0{=}0.55$, $\LR{=}6$).}
\label{tab:profiles}
\begin{tabular}{@{}llrr@{}}
\toprule
Profile & Mechanisms & Reliability & Calls \\
\midrule
Cheap & single solve & 0.55 & 1.0 \\
Balanced & vote\,5 + verify $k{=}2$ & 0.98 & 16.0 \\
High-reliability & verify $k{=}4$ (+abstain) & 0.999 & 17.3 \\
\bottomrule
\end{tabular}
\end{table}

\begin{table}[t]
\centering\small
\caption{Cost-to-target (simulation). Verification reaches each target
with slowly growing incremental cost; the voting-only column is the
approximate cost for a $p{=}0.62$ solver (normal approximation) and is
roughly an order of magnitude larger.}
\label{tab:target}
\begin{tabular}{@{}lrrr@{}}
\toprule
Target reliability & Config & Verify calls & Voting calls \\
\midrule
0.90  & $k{=}1$ & 3.8  & $\sim$27 \\
0.97  & $k{=}2$ & 7.4  & $\sim$60 \\
0.99  & $k{=}3$ & 11.7 & $\sim$90 \\
0.999 & $k{=}4$ & 17.3 & $\sim$160 \\
\bottomrule
\end{tabular}
\end{table}

\paragraph{Summary.}
The harness reproduces every qualitative prediction of the theory and
quantifies the trade-offs an operator faces: verification is the
high-leverage mechanism when a discriminating checker exists, voting is a
cheap supplement bounded by diversity, and abstention converts confidence
into reliability. The controller's marginal-rate policy reduces all of
these choices to two settings: a target reliability and a budget.

\section{Case Studies}\label{sec:cases}
The same harness specializes to very different tasks by changing only the
solver pool and the verifier ensemble; the controller and invariants are
unchanged. We sketch three configurations to make the mapping concrete.

\paragraph{Grade-school and competition mathematics.}
The base solver samples chain-of-thought solutions at moderate temperature;
diversity comes from multiple seeds and two prompt styles. The verifier
ensemble is deliberately \emph{robust}: a deterministic arithmetic
re-evaluator recomputes the final expression, a units and sanity
checker rejects out-of-range answers, and, for problems with a checkable form,
substitutes the candidate back into the problem's constraints.
Because these checks reduce to executing the problem's own definition,
their false-acceptance is low and stable, so $\LR$ is high and the
controller leans heavily on gating, falling back to self-consistency voting
only when no closed-form check applies. The expected behavior is the
amplification curve of Figure~\ref{fig:amp}: a few gates take a
near-threshold solver to high reliability at modest cost.

\paragraph{Program synthesis.}
Here verification is nearly ideal. The base solver proposes a program; the
verifier executes it in the sandbox against unit tests, a type-checker, and
property-based fuzzing, with failures counting as rejections
(invariant~I2). Execution-based checking has very high discrimination: a program that
passes a strong test suite is very likely correct, and one that fails is
certainly rejected. The controller's dominant mechanism is therefore a
single high-$\LR$ gate plus refinement: on rejection, the failing test
output is fed back as a critique and the solver revises. Decomposition
appears naturally for multi-function tasks, with each function gated by its
own tests before integration; this is the recursion master theorem
realized as test-driven assembly. This is the regime where orchestration most clearly
beats single-shot generation.

\paragraph{Open-domain question answering.}
Verification is hardest here because correctness is not mechanically
checkable. The harness leans on \emph{grounding}: a retrieval step supplies
sources, the solver must answer with citations, and the verifier checks
that each claim is supported by a retrieved passage (a corroboration check)
and cross-examines with an independent judge model prompted to find
unsupported assertions. No single check is decisive, so the controller
combines a modest verifier panel with self-consistency over diverse
retrievals, and, crucially, sets a conservative abstention threshold,
declining or escalating when corroboration is weak. The operating point
sits on the risk--coverage frontier of Figure~\ref{fig:rc}: the system
answers confidently where sources agree and abstains where they do not,
which is the responsible behavior for factual tasks.

\paragraph{Adaptive allocation versus fixed pipelines.}
A fixed pipeline, such as ``always sample five and verify twice,''
spends the
same budget on every instance, over-paying on easy cases and under-serving
hard ones. The controller instead reads each instance's running log-odds
and stops early when the target is already met, redirecting budget to the
instances that need it. Because difficulty is heavy-tailed (most instances
are easy, a few are hard), this adaptivity is where much of the practical
cost saving comes from: the average cost is dominated by the easy majority,
on which the controller commits quickly, while the hard tail receives the
deep verification it requires. The fixed pipeline is the special case the
controller reduces to when every instance happens to have the same
difficulty, which is rare in practice.

\section{Discussion: Failure Modes and Guidance}\label{sec:discussion}
\paragraph{Verifier gaming (Goodhart).}
When solvers are optimized against a fixed checker, the checker's effective
false-acceptance rises and its discrimination collapses toward
$\LR=1$, and amplification silently fails. (This is Goodhart's law: a proxy
measure stops tracking what it was meant to measure once it is optimized
directly.) Mitigations: prefer
\emph{robust} verifiers whose soundness is independent of the generator
(execution, proofs, types, corroborated retrieval); monitor each
verifier's online $\widehat\LR$ and demote drifters; rotate or ensemble
diverse verifiers; and hold out a ground-truth stream for recalibration.

\paragraph{Correlated errors.}
A pool of near-identical solvers inherits shared blind spots; voting then
plateaus far below $1$ (Fig.~\ref{fig:div}). Mitigation: engineer
diversity (distinct prompts, temperatures, tools, and ideally distinct
models) and measure $\gamma$ directly from disagreement rates.

\paragraph{Decomposition error compounding.}
Aggressive decomposition multiplies per-stage error and can underperform a
monolithic solve. Mitigation: decompose only when a clean split is
confident, and gate every recombination boundary so each level is restored
before it propagates; this is the recursion master theorem in
practice~\cite{companion}.

\paragraph{Cost and latency.}
Reliability is bought with calls; the controller's budget cap and
abstention keep total cost bounded, and caching plus parallel fan-out
keep latency manageable. The right operating point is application-specific and
is exactly the marginal-rate threshold $\lambda$.

\paragraph{When not to use a harness.}
Orchestration is not free, and not every task warrants it. For tasks where
the base solver is already reliable enough, where no informative verifier
exists ($\LR\approx1$), or where latency dominates and a single fast answer
is preferable to a slow reliable one, the cheap profile (single solve) is
correct and the overhead of voting and gating is waste. The controller
makes this an explicit decision: if the estimated marginal rate of every
mechanism is below $\lambda$ from the start, it commits the single solve.
The harness adds value exactly in the regime the theory identifies:
weak base solvers with discriminating, robust checkers.

\paragraph{Interaction with model improvements.}
A better base model raises $p_0$ and shifts the whole cost--reliability
curve left: the same target is reached with fewer gates and votes, and the
controller spends less by default. Crucially, orchestration and model
quality are complements, not substitutes: a stronger model with a
robust verifier reaches extreme reliability more cheaply than either
alone. The harness therefore remains valuable as models improve, simply
operating at a cheaper point. The exception is verification quality: if generation improves faster
than checking, the verifier becomes the limiting factor, which reinforces
the paper's central recommendation to invest there.

\paragraph{Hallucinations and long-running jobs.}
Two deployment problems map directly onto the harness. The first is
\emph{hallucination}: a generative model's confident production of
incorrect or unsupported content~\cite{ji2023hallucination}. This is
exactly the failure that calibrated confidence targets: a hallucinated
candidate either fails corroboration and test gates and is rejected, or
fails to accumulate enough log-odds and becomes an abstention or an
escalation. Confident errors are thereby converted into checked
refusals. The second is the \emph{long-running, multi-step job}: because
per-step error compounds, even a step that is $99\%$ reliable fails
almost surely over hundreds of steps. Verifying every boundary bounds
end-to-end error at logarithmic overhead (the recursion master theorem
of the companion theory), and idempotent, replayable state lets such
jobs checkpoint, suspend, and resume, which is what long-horizon
automation requires.

\paragraph{Safety.}
Fail-closed verification (treating errors and timeouts as rejections),
calibrated abstention, and full traceability make the harness conservative
by construction: when it is unsure, it declines or escalates rather than
emitting a confident guess. The full execution trace also makes every
answer auditable after the fact (which candidates were generated, which
verifiers accepted, and why the controller stopped), itself a
requirement for high-stakes deployment.

\section{Limitations and Threats to Validity}\label{sec:limits}
The headline results are \emph{simulations} of a parameterized model, not
measurements of a deployed system; they validate the harness logic and the
reliability laws but do not, by themselves, establish absolute numbers on
any benchmark. Real models violate the simulation's clean assumptions:
verifier errors are not perfectly conditionally independent, solver
correctness is instance-dependent rather than a single $p$, and $\beta,
\alpha,\gamma$ vary across the input distribution. The companion
theory~\cite{companion} bounds the cost of these violations (correlated
verifiers lose effective $\LR$; correlated solvers can hit a floor when
shared causes push the committee below chance), and the
harness estimates the relevant quantities online, but a full real-model
evaluation under the \S\ref{sec:eval} protocol is necessary to claim
deployment numbers and is the natural next step. Finally, our controller
is greedy; the constrained optimum it approximates can in principle be
solved more globally when marginal-rate estimates are reliable.

\section{Conclusion}\label{sec:conclusion}
\paragraph{Generalization.}
Although we describe the harness in terms of language models, nothing in
the architecture or the controller depends on that choice. A base solver is
any process behind the \texttt{solve} interface (a model, a heuristic, a
search procedure, a tool, a human, or another harness), and a verifier is
anything that returns a calibrated accept/reject. The same controller would
organize a pool of human reviewers with an automated checker, or a mix of
models and classical algorithms, identically. This is the practical payoff
of grounding the system in an algebra over abstract solvers: the harness is
a general-purpose organizer of unreliable problem-solving, of which the
language-model case is one instance.

We presented a model-agnostic harness that organizes unreliable solvers
into reliable problem-solving systems by compiling four structural
combinators into runtime components and a budget-aware controller that
spends compute where it buys the most reliability. A faithful simulation
reproduces the predicted laws: geometric amplification by verification,
threshold-and-floor behavior of voting, and the value of abstention. It
also shows that a controller-driven organization reaches reliability
targets at a small fraction of the cost of voting alone, automatically selecting
the cheapest mechanism for the regime at hand. The practical
message mirrors the theory: \emph{build robust checkers, diversify
solvers, verify every boundary, and let the controller allocate at the
margin}. Reliability is not a property of a model; it is a property of how
models are organized.

\section{Artifact and Reproducibility Statement}\label{sec:artifact}
All experimental results in this paper are Monte Carlo simulations of the
parameterized solver/verifier model of \S\ref{sec:eval}; no proprietary
models, datasets, or services are required to reproduce them. The full
source code is publicly available at
\url{https://github.com/hidayetaksu/maestro-order}, together
with pinned dependencies, reference outputs, and a README with
step-by-step instructions.

\paragraph{One-command reproduction.}
With Python~3.10+ and the pinned packages from
\texttt{requirements.txt} (NumPy~2.2.6 is the version that matters for
bit-exact randomness),
\begin{center}
\texttt{python reproduce.py all}
\end{center}
regenerates every artifact in two to four minutes on a single CPU core,
writing to \texttt{./output}. Table~\ref{tab:repro} maps each artifact to
its subcommand and seed. The four panels of
Figure~\ref{fig:results} and the ablation table share one pseudo-random
stream in a fixed order, so the \texttt{figures} subcommand runs them
together; the exact pattern of random draws inside the simulation
primitives is documented in the code and is part of the artifact.

\begin{table}[t]
\centering\small
\caption{Reproduction map: each published artifact, the subcommand of
\texttt{reproduce.py} that regenerates it, and the PCG64 seed.}
\label{tab:repro}
\begin{tabular}{@{}lll@{}}
\toprule
Artifact & Subcommand & Seed \\
\midrule
Fig.~\ref{fig:results}(a--d), Table~\ref{tab:ablation} & \texttt{figures} & 20260607 \\
Table~\ref{tab:trace} (trace, stop rate) & \texttt{trace} & 7 \\
Fig.~\ref{fig:traj} (trajectories) & \texttt{traj} & 11 \\
Closed-form numbers in the text & \texttt{predictions} & --- \\
\bottomrule
\end{tabular}
\end{table}

\paragraph{Verification.}
Reproduced outputs can be checked against the published artifacts by
direct comparison: the regenerated ablation table in \texttt{output/}
is byte-identical to the published copy in \texttt{figures/}, and JSON
summaries of every data series (including the measured maximum deviation
between simulation and the closed-form odds law, $0.0019\le0.004$) are
provided as reference outputs. Derived quantities quoted in the
text (the voting cost-to-target estimates of Table~\ref{tab:target},
the diversity plateaus of \S\ref{sec:results}, and the marginal rates in
the trajectory analysis) are computed in closed form by the
\texttt{predictions} subcommand, with no random input. The paper itself
builds from the repository root with \texttt{pdflatex} and the ACM
\texttt{acmart} class; figure and table sources are read from the
\texttt{figures/} directory that the reproduction script can regenerate.

\end{document}